\documentclass[superscriptaddress,prl,twocolumn,tightenlines,balancelastpage,10pt,a4paper]{revtex4}%
\usepackage{bbm}
\usepackage{amsfonts}
\usepackage{amssymb}
\usepackage{graphicx}
\usepackage{subfigure}
\usepackage{savesym}
\usepackage{amsmath}
\usepackage{txfonts}
\usepackage{multirow}
\usepackage{epstopdf}

\usepackage{soul}

\usepackage{color}

 \savesymbol{iint}
\restoresymbol{TXF}{iint}

\newcommand{\removed}[1] {\textcolor[gray]{0.7}{#1}}
\newcommand{\changeda}[1] {\textcolor{black}{#1}}
\newcommand{\changedb}[1] {\textcolor{black}{#1}}

\newcommand{\changede}[1] {\textcolor{black}{#1}}

\begin{document}

\author{Xiao-Qi Zhou}
\affiliation{Centre for Quantum Photonics, H. H. Wills Physics
Laboratory \& Department of Electrical and Electronic Engineering,
University of Bristol, BS8 1UB, United Kingdom}
\author{Pruet Kalasuwan}
\affiliation{Centre for Quantum Photonics, H. H. Wills Physics
Laboratory \& Department of Electrical and Electronic Engineering,
University of Bristol, BS8 1UB, United Kingdom}
\author{Timothy C. Ralph}
\affiliation{Centre for Quantum Computation and Communication Technology, School of Mathematics and Physics, University of Queensland, Brisbane 4072, Australia}
\author{Jeremy L. O'Brien}
\email{Jeremy.OBrien@bristol.ac.uk}
\affiliation{Centre for Quantum Photonics, H. H. Wills Physics
Laboratory \& Department of Electrical and Electronic Engineering,
University of Bristol, BS8 1UB, United Kingdom}

\title{{{\color{black}{Calculating Unknown Eigenvalues with a Quantum Algorithm}}}\vspace{-6pt}}

\begin{abstract}
Quantum algorithms are able to solve particular problems exponentially faster than conventional algorithms, when implemented on a quantum computer. However, all demonstrations to date have required already knowing the answer to construct the algorithm. We have implemented the complete quantum phase estimation algorithm for a single qubit unitary in which the answer is calculated by the algorithm. We use a new approach to implementing the controlled-unitary operations that lie at the heart of the majority of quantum algorithms that is more efficient and does not require the eigenvalues of the unitary to be known.  These results point the way to efficient quantum simulations and quantum metrology applications in the near term, and to factoring large numbers in the longer term. This approach is architecture independent and thus can be used in other physical implementations.
\end{abstract}

\maketitle \noindent

A quantum algorithm is a program, designed to be run on a quantum computer, that solves a computational task using less physical resources than the best known classical algorithm \cite{nielsen}; of most interest are those for which an exponential reduction 
is achieved. \changeda{The} key example is the phase estimation algorithm \cite{ki-eccc-3-3}, which \changeda{provides the quantum speedup in} Shor's factoring algorithm \cite{sh-conf-94-124} and quantum simulation algorithms \cite{ll-sci-273-1073,as-sci-309-1704}.
\changeda{To date experiments have demonstrated only} the read-out phase of \changeda{quantum} algorithms, 
but not the steps in which input data is read-in and processed in order to calculate the final quantum state; 
knowing the answer beforehand was essential in preparing this state. 

We present a full\changeda{, scalable} demonstration of the \changeda{iterative} quantum phase estimation algorithm~\changeda{(IPEA)} for a one-qubit unitary in which the extracted answer is truly calculated by the algorithm. 
In contrast to previous work
, no prior knowledge of the unitary is required for the implementation of the algorithm.
We also use the IPEA circuit to generate eigenstates of a unitary, which is important in the context of quantum chemistry simulation algorithms \cite{ka-arpc-625-185}, for example. \changede{The scheme itself is architecture independent and can be used in other physical architectures.}
These results point to practical applications of the phase estimation algorithm, including quantum simulations and quantum metrology in the near term, and factoring in the long term.

Quantum computation---the processing and read-out of information encoded on quantum systems---can in principle solve computational problems that are intractable using classical computation. Examples of classically intractable problems that can be solved by a quantum computer are the determination of the prime factors of a large number and the simulation of complex chemical reactions. In both these examples, the key quantum sub-routine employed is the phase estimation algorithm which, given some non-diagonal representation of a Hamiltonian plus one of its eigenstates, enables the corresponding eigenvalue to be estimated to arbitrary precision in polynomial time.

Many quantum computations can be roughly broken down into two sections: read-in and processing of the input data; and processing and read-out of the solution. In the first phase the initial data is read in to a quantum register and processed with quantum gates, sometimes multiple times. This produces a quantum state in which the solution is encoded. In the second phase the quantum state may be subjected to further processing followed by measurement, producing a classical data string containing the solution. Even though quantum computers are currently limited to a small number of qubits, there is considerable interest in the small scale demonstration of quantum algorithms even if the size of the problems solved means that they remain easily tractable with classical techniques. Such demonstrations remain challenging even for small numbers of qubits as they typically require the sequential application of a large number of quantum gates.

In recent years a number of elegant demonstrations of the read-out phase of Shor's factoring algorithm \cite{va-nat-414-883,lu-prl-99-250504,la-prl-99-250505,po-sci-325-1221} and a quantum chemistry simulation algorithm \cite{la-nchem-2-106} have been made. In these demonstrations, quantum gates have been used to produce the quantum state corresponding to a particular solution of the algorithm. It was then shown that the corresponding solution could be read-out with high fidelity from this state. However, in each case, the method for producing the quantum state explicitly required the solution to already be known from a classical calculation. That is, the solution was put into the quantum state by hand, before being read-out through further processing and measurement. It is clearly important to go beyond this restriction and demonstrate both phases of a quantum algorithm. 



\begin{figure}
\vspace{-0.5cm}
\includegraphics[width=.5\textwidth]{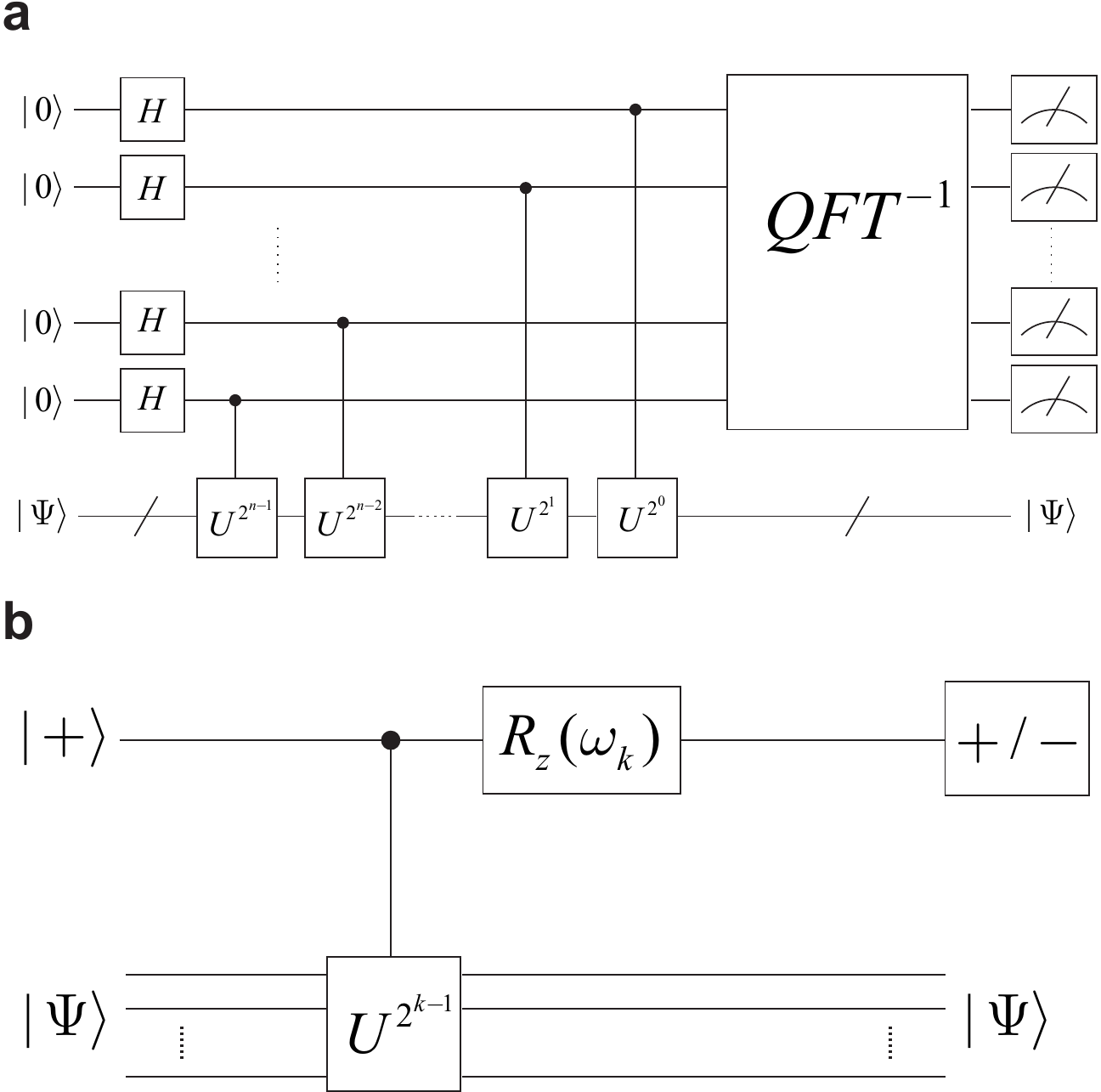}
\caption{The phase estimation algorithm. (a) The standard quantum circuit for phase estimation, where $H$ is the standard Hadamard gate and $QFT^{-1}$ is the inverse quantum Fourier transform. The measurements are all implemented in computational basis. (b) The $k^{th}$ iteration of the iterative phase estimation algorithm~(IPEA). The algorithm is iterated $m$ times to get an $m$-bit $\tilde{\varphi}$~(see Eq.~\ref{eq2}), which is the approximation to the phase of the eigenstate $\varphi$~(see Eq.~\ref{eq1}). \changeda{The measurement is implemented in $+/-$ basis, where $|+$/$-\rangle=\frac{1}{\sqrt{2}}\left(|0\rangle\pm|1\rangle\right)$.} Each iteration obtains one estimated bit $\tilde{\varphi}_k$; starting from the least significant~($\tilde{\varphi}_m$), $k$ is iterated backwards from $m$ to 1. The feedback angle $\omega_k$ depends on the previously measured bits as $\omega_k=-2\pi\xi_k$, where $\xi_k=0.0\tilde{\varphi}_{k+1}\tilde{\varphi}_{k+2}...\tilde{\varphi}_{m}$ in binary expansion and $\omega_m=0$.
}
\label{fig1}
\vspace{-0.5cm}
\end{figure}

First, we briefly review the standard phase estimation algorithm~\cite{nielsen}: Given a unitary $U$ and one of its eigenstates $\vert \psi \rangle$ which fulfill the equation
\begin{equation}
U\vert \psi \rangle=e^{i2\pi\varphi}\vert \psi \rangle
\label{eq1}
\end{equation}
the task is to find what the corresponding eigenvalue is---in other words, find the value of $\varphi$. 
As shown in Fig.~\ref{fig1}(a), $m$ ancillary qubits act as controls, where each qubit is prepared in $\vert 0\rangle$, and the target is the given eigenstate $\vert \psi \rangle$. After applying a Hadamard gate to each of the control qubits, we obtain the state $\vert +\rangle^{\otimes m} \otimes \vert \psi \rangle$, where $\vert +\rangle=\frac{1}{\sqrt{2}}\left(\vert 0\rangle+\vert 1\rangle\right)$. This state can also be represented as
\begin{equation}
\sum_{x=0}^{2^{m+1}-1}{\vert x\rangle}\otimes \vert \psi \rangle.
\end{equation}
Then a series of controlled-unitary gates are applied on the state as shown in Fig.~\ref{fig1}(a) and thus convert it to
\begin{equation}
\sum_{x=0}^{2^{m+1}-1}{\vert x\rangle\otimes U^{x}\vert \psi \rangle} = \left(\sum_{x=0}^{2^{m+1}-1}{e^{i2\pi \varphi x}\vert x\rangle}\right)\otimes \vert \psi \rangle.
\end{equation}
The target state is intact and all the information about $\varphi$ is contained in the state of the control qubits.
The $m$ qubits of the control register then undergo an inverse quantum Fourier transform~($QFT^{-1}$), and the control qubits are converted to $\vert \tilde{\varphi}_1 \rangle\otimes\vert \tilde{\varphi}_2 \rangle...\otimes\vert \tilde{\varphi}_m \rangle$, where $\tilde{\varphi}_i$~($1\leqslant i \leqslant m$) is an estimated bit equal to 0 or 1. By measuring the control qubits in the computational basis, one obtains the values of $\tilde{\varphi}_1$, $\tilde{\varphi}_2$...$\tilde{\varphi}_m$ and the estimated phase in binary expansion:
\begin{equation}
\tilde{\varphi}=0.\tilde{\varphi}_1\tilde{\varphi}_2...\tilde{\varphi}_m
\label{eq2}
\end{equation}

As the inverse quantum Fourier transform can be \changeda{scalably} realized in a semiclassical way~\cite{gr-prl-76-3228} where no entangling gates are needed, 
the circuit with $m$ ancillary qubits in Fig.~\ref{fig1}(a) can be simplified to an $m$ round iterative single ancillary qubit circuit. This simplified version is called the iterative phase estimation algorithm (IPEA)~\cite{mi-pra-76-030306}. Figure~\ref{fig1}(b) shows the IPEA at the $k^{th}$ iteration. %
At the end of this iteration, a measurement of the ancillary qubit in the computational basis is performed, yielding the result 0 or 1, which is the estimate of the $k^{th}$ bit of $\varphi$ in the binary expansion. Note that in the IPEA scheme the least significant bits are evaluated first (that is, k is iterated backwards from m to 1) and the information obtained is used to improve the estimation of the more significant bits. This information transfer between iterations is realized via a single qubit rotation $R_z(\omega_k)$, whose angle is determined by all previously measured bits, as described in the caption of Fig.~\ref{fig1}(b).

\begin{figure}
\vspace{-0.5cm}
\includegraphics[width=.5\textwidth]{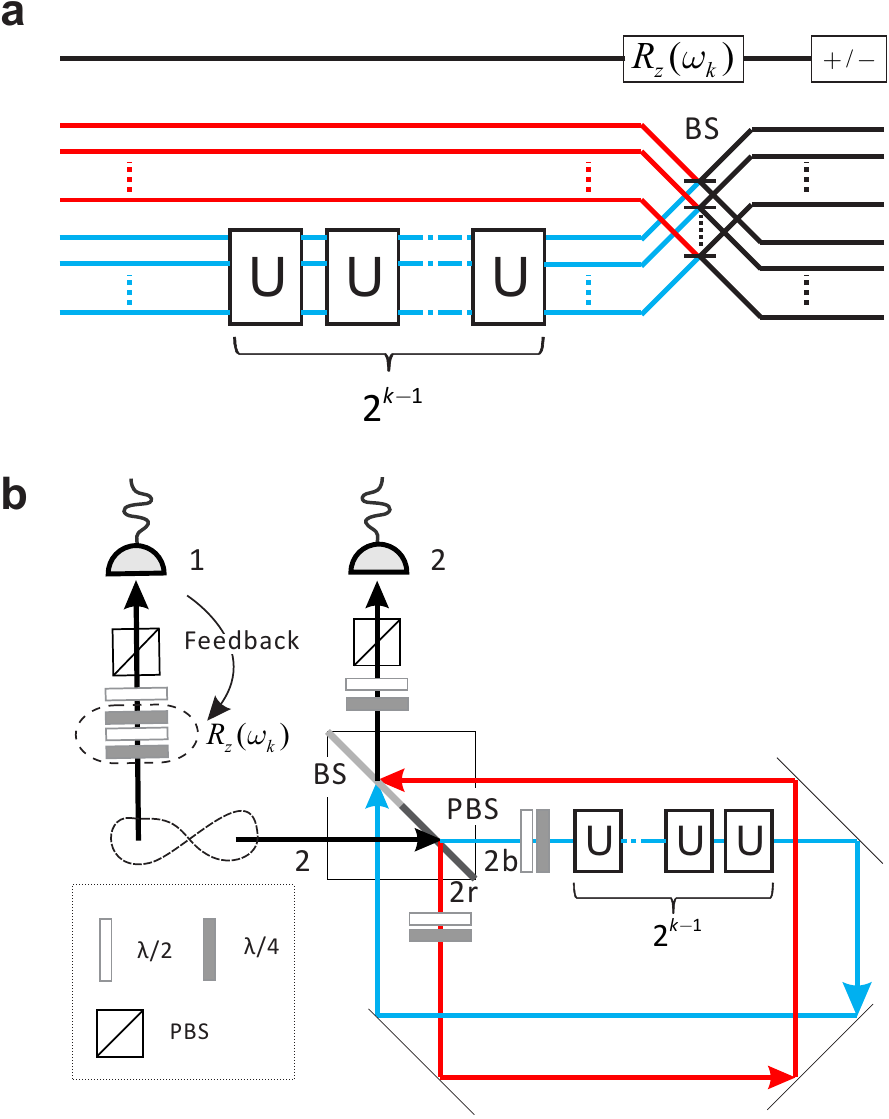}
\vspace{-0.5cm}
\caption{
\changede{(a)} Simplified entanglement-based circuit for $C$-$U^{2^{k-1}}$ gate. The initial input state is \changeda{$\frac{1}{\sqrt{2}}\left(\vert H \rangle\otimes\vert \psi \rangle_{r} +\vert V \rangle \otimes \vert \psi \rangle_{b}\right)$}, where $\vert \psi \rangle$ is a multi-qubit polarization-encoded state
and red $r$ and blue $b$  denote different spatial modes of the photons.
After the blue mode\changeda{s} pass through the unitary gate $U^{2^{k-1}}$, which is realized by cascading $2^{k-1}$ copies of $U$, \changeda{the red and blue modes of each target qubit are mixed on beamsplitters (BS)}. By post-selecting the case where \changeda{an even number of target photons arrives in lower spatial modes}, $C$-$U^{2^{k-1}}$ is 
realized for the input state \changeda{$\vert +\rangle\otimes\vert \psi \rangle$}, where $\vert +\rangle=\frac{1}{\sqrt{2}}(\vert H \rangle + \vert V \rangle)$. \changeda{The rotation $R_z(\omega_k)$ and the measurement in $+/-$ basis are then used to extract $\tilde{\varphi_k}$---the estimate of the $k^{th}$ bit of the phase $\varphi$. \changede{(b) Experimental setup for the $k$th iteration of the two-qubit iterative phase estimation algorithm~(IPEA). A 60 mW continuous-wave (CW) laser beam with a central wavelength of 404nm is focused onto a type-II BBO crystal to create the polarization entangled photon-pairs. The PBS part of the BS/PBS cube and the following waveplates convert the two photons to the desired polarization-spatial entangled state~(see Equation~\ref{eq3}). Based on this state, the $C$-$U^{2^{k-1}}$ gate is effectively realized, where $U$ is the unitary whose eigenvalue is target to be estimated. 
The rotation gate $R_z(\omega_k)$~(for the value of $\omega_k$, see the caption of Figure~\ref{fig1}b) is implemented by three waveplates --- two quarter-waveplates with a half-waveplate in between. The displaced-Sagnac structure make the phase between modes $2r$ and $2b$ inherently stable.} }
}
\label{fig2}
\vspace{-0.5cm}
\end{figure}


In the case where the phase $\varphi$ has exactly $m$ bits in binary expansion, a perfect implementation of $m$ iterations of the algorithm will deterministically extract the exact phase, which means $\tilde{\varphi}=\varphi$. When $\varphi$ has a binary expansion more than m bits, it has been proven that a perfect implementation of the algorithm achieves a precision of $\pm2^{-m}$ with an error probability less than $19\%$, which is independent of $m$~\cite{mi-pra-76-030306}. This error can always be eliminated by simply repeating each IPEA iteration several times and choosing the most frequently observed result as the corresponding estimated bit. \changeda{Note that this procedure is scalable.}





From the description of the IPEA above, it is clear that the key procedure of this algorithm is to implement a sequence of controlled unitary gates 
$C$-$U^{2^{k-1}}$ (see Fig.~\ref{fig1}(b)). The standard method for implementing a controlled-unitary gate relies on its decomposition into controlled-NOT (CNOT) and single-qubit gates \cite{nielsen}. \changeda{Even f}or the simplest two-qubit case, where the unitary is a single-qubit unitary, two CNOT gates and three single-qubit gates are needed to realise this controlled-unitary gate~\cite{nielsen}. 

Recently, a scheme for simplifying the construction of controlled-unitary gates was proposed~\cite{ra-pra-75-022313} and experimentally demonstrated~\cite{la-nphys-5-134}. 
Based on this scheme, the realization of the \changeda{IPEA} in linear optics has been reported~\cite{la-nchem-2-106}.  However, this realization has a critical drawback: The method of constructing the controlled-unitary gate in that experiment is based on decomposing the single-qubit unitary $U$ to the product of $T$, \changeda{$R_z(\alpha)$} and $T^{-1}$, where $T$ and $T^{-1}$ are two complementary unitary gates and \changeda{$R_z(\alpha)$} is a phase-shift gate with \changeda{$\alpha$} phase shift in the computational basis. The paradox is that one can directly calculate the phase introduced by $U$ from the information of how to decompose $U$ in this way---\textit{i.e.} there is no need to use the phase estimation algorithm to estimate the phase because it is already known exactly. To overcome this and realize the phase estimation algorithm 
generally, 
control qubits should be added to the unitary without already knowing this information.

\begin{figure*}
\vspace{-0.5cm}
\includegraphics[width=\textwidth]{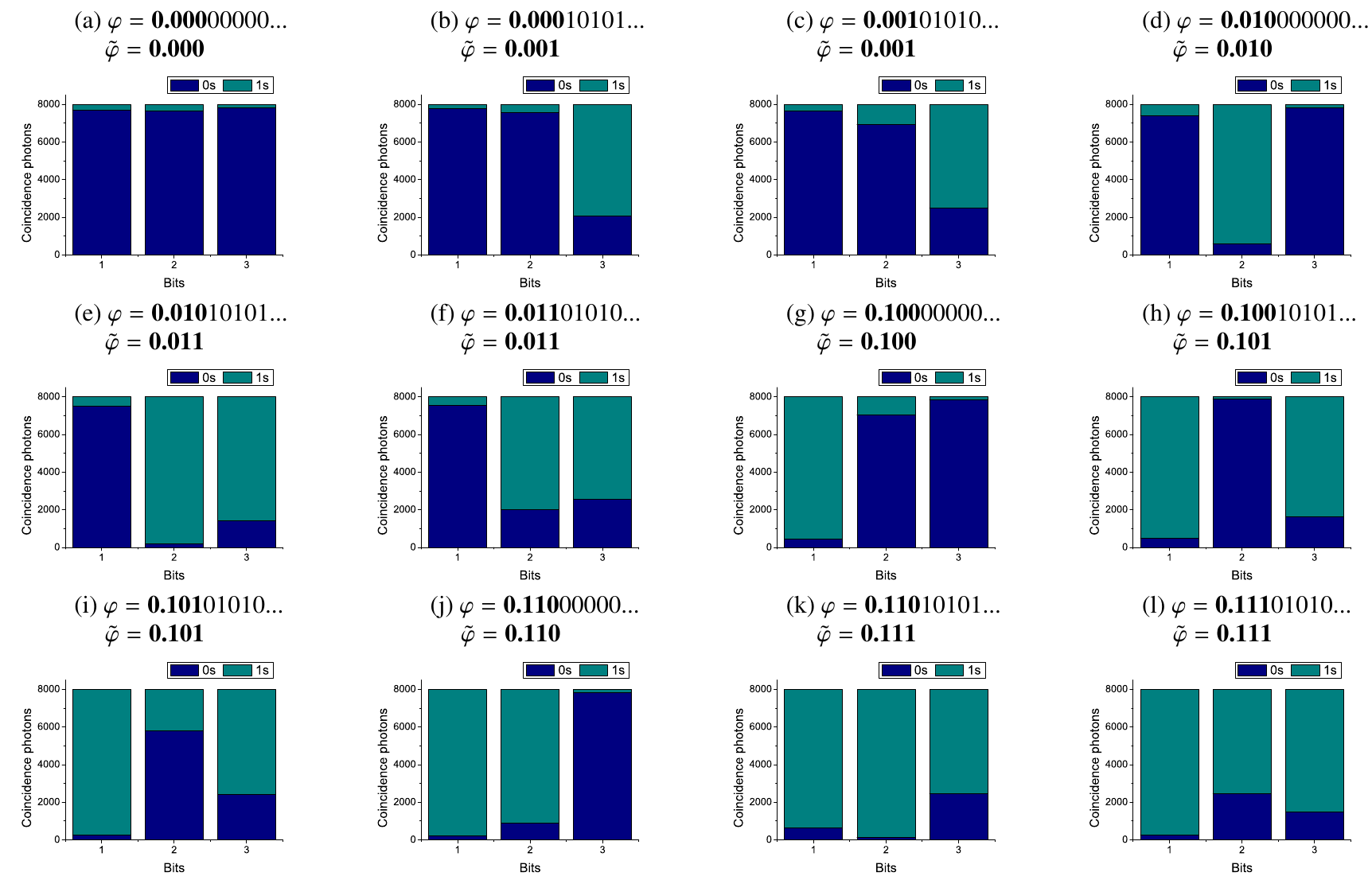}
\caption{Phase estimation data for 12 different $U$s. Each $U$ is composed of two half-waveplates~(HWPs); the first is set to $0^{\circ}$, the second HWP is oriented at (a) $0^{\circ}$, (b) $15^{\circ}$, (c) $30^{\circ}$, (d) $45^{\circ}$, (e) $60^{\circ}$, (f) $75^{\circ}$, (g) $90^{\circ}$, (h) $105^{\circ}$, (i) $120^{\circ}$, (j) $135^{\circ}$, (k) $150^{\circ}$ and (l) $165^{\circ}$. 
For each $U$, three iterations of the algorithm are implemented and thus a three-digits estimated phase $\tilde{\varphi}$ is obtained. Compared with the phase $\varphi$ the error in $\tilde{\varphi}$ is always less then $0.0001$ in binary, which is consistent with theoretical prediction.
}
\label{fig4}
\vspace{-0.5cm}
\end{figure*}

\begin{figure*}
\vspace{-0.5cm}
\includegraphics[width=\textwidth]{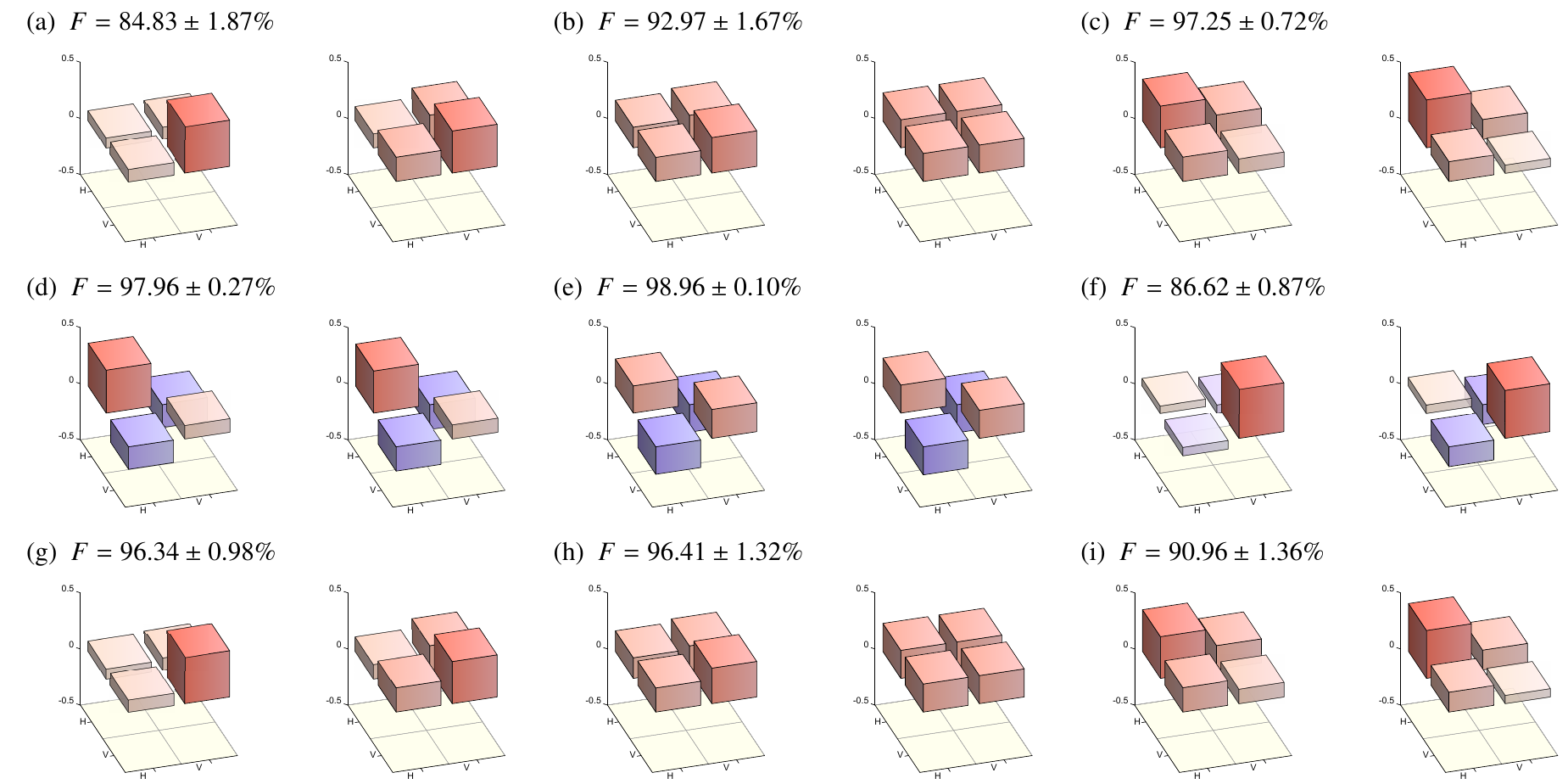}
\caption{Using the phase estimation algorithm to generate the eigenstates of $U$. Panels (a)-(i) show the density matrix of the target output; (L) experimental (R) ideal. Unitaries  $U_1$, $U_2$ and $U_3$ are implemented with a single half-waveplate~(HWP)set to $30^{\circ}$, $45^{\circ}$ and $67.5^{\circ}$, respectively. Panels (a), (b), (c) show the case where the unitary is $U_1$, $U_2$, $U_3$, the initial target state is $\vert H\rangle$ and the measurement outcome of the control qubit is $0$. Panels (d), (e), (f) show the case where the unitary is $U_1$, $U_2$, $U_3$, the initial target state is $\vert H\rangle$ and the measurement result of the control qubit is $1$. Panels (g), (h), (i) show the case where the unitary is $U_1$, $U_2$, $U_3$, the initial target state is $\vert V\rangle$ and the measurement result of the control qubit is $0$. The output target state is determined only by the measurement result of the control qubit and not affected by changing the initial target state, as verified by the similarity between the matrices (a) and (g), (b) and (h), (c) and (i). State tomography and maximum-likelihood are used for the reconstruction of the density matrices. The fidelities of the reconstructed density matrices with the ideal case are shown. The error estimates are obtained by performing many reconstructions with random noise added to the raw data in each case.
}
\label{fig5}
\vspace{-0.5cm}
\end{figure*}


We recently proposed and demonstrated a new scheme for adding 
control qubits to quantum gates~\cite{xq-ncomm-2-413}. In the case of the phase estimation algorithm, the key advantage of this scheme is that a controlled-unitary gate can be implemented without needing to know any information about the unitary---\textit{i.e.} the unitary can be a ``black box" and still the controlled-unitary can be implemented.
This feature is critical for the phase estimation application. Details of this scheme can be found in Ref.~\onlinecite{xq-ncomm-2-413};


\changeda{For the IPEA application, where the control qubit is always prepared in $\vert +\rangle$ state, the implementation of the controlled-unitary gate can be simplified.}  As shown in Fig.~\ref{fig2}\changede{(a)}, the initial state is \changeda{$\frac{1}{\sqrt{2}}\left(\vert H \rangle\otimes\vert \psi \rangle_{r} +\vert V \rangle \otimes \vert \psi \rangle_{b}\right)$, where $\vert H \rangle$ and $\vert V \rangle$ denote horizontal and vertical polarization respectively, $\vert \psi \rangle$ denotes the given (eigen)state encoded in $n$ polarization qubits, $r$ and $b$ denote the red and blue spatial modes respectively.}

\changeda{
The blue modes of the target pass through the unitary $U^{2^{k-1}}$ and thus the state is converted to
\begin{equation}
\frac{1}{\sqrt{2}}\left(\vert H \rangle\otimes\vert \psi \rangle_{r} +\vert V \rangle \otimes U^{2^{k-1}}\vert \psi \rangle_{b}\right)
\end{equation}
Then the red and blue modes of each target qubit are mixed on non-polarising beamsplitters (BS) to remove the path information and the state is now changed to
\begin{equation}
\begin{split}
& \sum\limits_{p \in P}\frac{1}{\sqrt{2^{n+1}}}\left(\vert H \rangle\otimes\vert \psi \rangle_{p}+ \vert V \rangle \otimes U^{2^{k-1}}\vert \psi \rangle_{p}\right)\\
+& \sum\limits_{q \in Q}\frac{1}{\sqrt{2^{n+1}}}\left(\vert H \rangle\otimes\vert \psi \rangle_{q}- \vert V \rangle \otimes U^{2^{k-1}}\vert \psi \rangle_{q}\right)
\end{split}
\end{equation}
where $P$~($Q$) denotes the cases where an even~(odd) number of target photons arrive in the lower spatial modes. By post-selecting any case in $P$, the desired state $\frac{1}{\sqrt{2}}\left(\vert H \rangle\otimes\vert \psi \rangle+ \vert V \rangle \otimes U^{2^{k-1}}\vert \psi \rangle\right)$ is obtained with a $\left(1/2\right)^n$  probability of success. There are $2^{n-1}$ such cases in $P$ that the total probability of success is $1/2$, regardless of the size of the unitary gate $U$.
}
Here $U^{2^{k-1}}$ is implemented by simply placing $2^{k-1}$ copies of the unitary $U$ into the path of the blue mode\changeda{s ($U^{2^{k-1}}$ could alternatively be realised by $2^{k-1}$ passes through $U$ \cite{hi-nat-450-393})}.
\changeda{
Finally, to finish the iteration, the first qubit passes through the rotation $R_z(\omega_k)$ and is measured in $+/-$ basis to extract $\tilde{\varphi_k}$ --- the estimate $k^{th}$ bit of the phase $\varphi$.
}
\changeda{
There is another $1/2$ probability that one of the cases in $Q$ occurs which means the state $\frac{1}{\sqrt{2}}\left(\vert H \rangle\otimes\vert \psi \rangle- \vert V \rangle \otimes U^{2^{k-1}}\vert \psi \rangle\right)$ is obtained. Using the same procedures will extract the same $\tilde{\varphi_k}$ as long as the measurement result of $+$~($-$) is redefined as $1$~($0$). In this way, the circuit shown in Fig.~\ref{fig2}\changede{(a)} can be used to implement the IPEA deterministically---\emph{i.e} with probability 1.
}




By using the entanglement-based controlled-unitary gates described above, we implement the IPEA without having to already know the value of the  phase $\varphi$---\textit{i.e.} without already knowing the answer to the algorithm. 
The experimental setup is shown in Fig.~\changede{\ref{fig2}(b)}: A $60$ mw 404 nm continuous wave laser is focused on a BBO crystal cut for type-II spontaneous parametric down-conversion (SPDC) to create a two-photon polarization-entangled state $\frac{1}{\sqrt{2}}\left(\vert H \rangle_1\otimes\vert V \rangle_{2}+\vert V \rangle_1\otimes\vert H \rangle_{2}\right)$, where $1$ and $2$ denote the control and target photons, respectively. A special beamsplitter cube, which on one half is a non-polarization beamsplitter~(BS) and the other half is polarization beamsplitter~(PBS) \cite{gao-2008}, is used as shown to build a displaced-Sagnac structure to increase the inherent phase stability of the setup. The photon $2$ passes through the PBS part of the BS/PBS cube and thus the two-photon state is converted to $\frac{1}{\sqrt{2}}\left(\vert H \rangle_1\otimes\vert V \rangle_{2r}+\vert V \rangle_1\otimes\vert H \rangle_{2b}\right)$. 
Waveplates are used in the path of $2r$ and $2b$ to prepare the required polarization-spatial entangled state
\begin{equation}
\frac{1}{\sqrt{2}}\left(\vert H \rangle_1\otimes\vert \psi \rangle_{2r}+\vert V \rangle_1\otimes\vert \psi \rangle_{2b}\right),
\label{eq3}
\end{equation}
where $\vert \psi\rangle$ is the eigenstate of the target unitary $U$---\textit{i.e} $U\vert \psi \rangle=e^{i\varphi}\vert \psi \rangle$. Then, after the blue mode passes through the unitary $U^{2^{k-1}}$, the two modes of photon 2 are combined at the BS side of the BS/PBS cube (Fig.\changede{~\ref{fig2}(b)}). By post-selecting the case where photon 2 exits at port 2, we get the desired two-photon state $\frac{1}{\sqrt{2}}\left(\vert H \rangle_1\otimes\vert \psi \rangle_{2} +\vert V \rangle_1 \otimes U^{2^{k-1}}\vert \psi \rangle_{2}\right)$, which can be written as $\frac{1}{\sqrt{2}}\left(\vert H \rangle_1 +e^{i2\pi\varphi2^{k-1}}\vert V \rangle_1 \right)\otimes\vert \psi \rangle_{2}$. To finish the $k^{th}$ iteration, photon 1 passes through the $R_z(\omega_k)$ gate~($\omega_k$ is set to an angle determined by all previously measured bits; see Fig.~\ref{fig1}(b) caption) and then is measured in the \changeda{$+/-$} basis to obtain the $k^{th}$ bit of the estimated phase.







We implemented three iterations of the IPEA to estimate the value of the phase $\varphi$ to three bits of precision. The unitaries $U^4$, $U^2$, and $U$ are used in the first, second and third iterations, and $U^4$ and $U^2$ are realized by four and two consecutive $U$ gates, respectively. 
The $U$ gate is implemented by two consecutive HWPs. A convenient feature of this unitary is that $\vert R\rangle$ and $\vert L\rangle$ are always eigenstates, where $\vert R/L\rangle=\frac{1}{\sqrt{2}}\left(\vert H \rangle\pm i\vert V \rangle\right)$. This can be understood by considering the following fact: a HWP always convert the states $\vert R\rangle\leftrightarrow\vert L\rangle$ no matter what the angle of the HWP is. Two consecutive HWPs therefore leave $\vert R\rangle$ and $\vert L\rangle$ unchanged, up to a phase factor---\textit{i.e.} $\vert R\rangle$  and $\vert L\rangle$ are the eigenstates of this unitary $U$. We therefore choose $\vert R\rangle$ as the input eigenstate $\vert \psi\rangle$. Other than convenience there is nothing special about this choice of $U$. The eigenvalue of $U$ is determined by the angles of the two HWPs. 
We fixed the angle of the first HWP to $0^\circ$ and changed the second HWP's angle $\theta$ to various values to realize a number of different unitaries. For each of these unitaries, we get a 3-digit estimate of the phase $\tilde{\varphi}$ in binary expansion. The results are shown in Fig.~\ref{fig4}. We see that the IPEA gives an accurate estimate of the phase $\varphi$. We note: (i) that $\varphi$ is non-trivially related to $\theta$, i.e. $\theta$ appears in a non-diagonal representation of U which must be diagonalized to extract $\varphi$; and (ii) in principle a third party could prepare the waveplates without revealing $\theta$, and the experimenter would still be able to successfully extract $\varphi$.





It has been shown \cite{da-prl-83-5162} that the phase estimation algorithm still works even when the input target state is not the eigenstate of $U$~(provided the iterations are coherent): 
Assume the input state is $\alpha \vert \psi_a \rangle+\beta \vert \psi_b \rangle$ where $\vert\psi_a \rangle$ and $\vert\psi_b \rangle$ are the eigenstates of $U$ with distinctive eigenvalues $e^{2\pi \varphi_a}$ and $e^{2\pi \varphi_b}$ respectively. By passing the control and the target through the same circuit as shown in Fig.~\ref{fig1}(a), the state $\alpha \vert \tilde{\varphi}_a \rangle \otimes \vert \psi_a \rangle+\beta \vert \tilde{\varphi}_b \rangle \otimes \vert \psi_b \rangle$ would be obtained at the output, where $\tilde{\varphi}_a$ and $\tilde{\varphi}_b$ are the estimates of $\varphi_a$ and $\varphi_b$, respectively. When the number of control qubits is sufficiently large to make $\tilde{\varphi}_a$ and $\tilde{\varphi}_b$ distinguishable, measuring the control qubits in the computational basis yields either $\tilde{\varphi}_a$, which means the estimated eigenvalue is $e^{2\pi \tilde{\varphi}_a}$ and the target state automatically collapses to the corresponding eigenstate $\vert \psi_a \rangle$, or $\tilde{\varphi}_b$, which means the estimated eigenvalue is $e^{2\pi \tilde{\varphi}_b}$ and the target state collapses to the state $\vert \psi_b \rangle$. In this way, the phase estimation circuit can be regarded as an eigenvalue measuring device or as an eigenstate generator.

%
We performed an experiment to show this eigenstate generation feature of the phase estimation algorithm. We use \changeda{a} similar experimental setup as shown in Fig.~\changede{\ref{fig2}(b)}. The $U$ gate, whose eigenvalue is the target to be estimated, is implemented by a single HWP. Two HWPs oriented at the same angle realize an identity operator, which means $U^2=I$. Based on this fact, one can deduce that, for $k\geq2$, all the gates $C$-$U^{2^{k-1}}$ are equal to identity operators. No iterations are required and we implement the only non-trivial circuit $C$-$U$~(corresponds to $k=1$). 
We set the initial target state to $\vert H \rangle$ and measure the control photon in the $+/-$ basis. When the result is $+$ or $-$, the target qubit collapses to the eigenstate of $U$ with eigenvalue $+1$ or $-1$, respectively. To evaluate this process, we perform state tomography on the output target state and compare the result with the theoretical prediction. The results are shown in Figs.~\ref{fig5}(a)--(f). The output target state, which is only determined by the measurement result of the control qubit, is not affected by changing the initial target state. We verify this feature by changing the initial target state to $\vert V \rangle$ and performing the state tomography on the output target state. The results are shown in Figs.~\ref{fig5}(g)--(i). 

The non-unit fidelities observed above arise primarily due to two effects: the partial distinguishability of the photons generated in the SPDC source that results in some incoherent mixture; and the imperfect optical components, including settings of waveplate angles. In the case of the eigenstate generation results (Fig.~\ref{fig5}), the fidelities of the output states range from $84.83\%$ to $98.96\%$
, which is, in part, due to the varying overlap between the input state $\vert H \rangle$ and the output states
; this results in a large difference in the output count rates. The total counts of the output are $\sim$600 s$^{-1}$ in~Fig.~\ref{fig5}(a) and $\sim$2200 s$^{-1}$ in Fig.~\ref{fig5}(c), for example, while the error counts of the output are similar in both cases. 

The ability to construct quantum algorithms without the need to know the answer in advance is clearly essential to their practical application. In the case of the phase estimation algorithm
this requires that the controlled unitaries $C$-$U^{2^{k-1}}$ are realised without already knowing the eigenvalues of $U$. The approach demonstrated here achieves this \changedb{efficiently} and opens the way to practical applications of quantum simulation algorithms, for calculating molecular properties~\changede{\cite{la-nchem-2-106}} for example, and metrology applications~\cite{hi-nat-450-393}, for enhanced measurement precision, in the near term, and factoring in the long term. \changede{For Shor's factoring algorithm, coherent iterations are required~\cite{pa-prl-85-3049}, which can be realized by employing the path entangling gates of Ref.~\onlinecite{xq-ncomm-2-413}.}

We note that the approach taken here is scalable:
the IPEA is itself scalable;
the scheme of Fig.~\ref{fig2}\changede{(a)} works deterministically, provided $U$ can be performed and the entangled input state can be prepared;
the realization of controlled unitaries is scalable to multi-qubit unitaries and applicable to any physical implementation where a higher dimensional Hilbert space is accessible, which typically is the case. Trapped ions systems, for example, offer a large number of precisely controllable internal electronic and external vibrational degrees of freedom.
For photons path degrees of freedom, as used here, are ideal, and the required entangled input state can be efficiently prepared using the `KLM' \cite{kn-nat-409-46} or derivative approaches to linear optical quantum computing. For other architectures this entangled input may be directly prepared using deterministic entangling gates, and the BS operations (Fig.~\ref{fig2}\changede{(a)}) would correspond to single qudit operations. 

\vspace{6pt}
\noindent We thank P.J. Shadbolt for writing the quantum process tomography code 
and J.C.F. Matthews,  A. Peruzzo, G.J. Pryde and P. Zhang for helpful discussions. This work was supported by EPSRC, ERC, PHORBITECH and NSQI. J.L.O'B. acknowledges a Royal Society Wolfson Merit Award.

\end{document}